\documentclass[prb,twocolumn,showpacs,superscriptaddress]{revtex4}

\usepackage{graphicx}
\usepackage{amsmath}
\usepackage{amssymb}
\usepackage{epsfig}

\renewcommand{\v}[1]{{\bf #1}}
\newcommand{\h}[1]{{\hat #1}}

\newcommand{\w}{{\omega}}

\def\eqa{\begin{eqnarray}}
\def\eea{\end{eqnarray}}
\newcommand{\eq}{\begin{equation}}
\newcommand{\ee}{\end{equation}}
\newcommand{\nn}{\nonumber\\}

\newcommand{\<}{\langle}
\renewcommand{\>}{\rangle}
\newcommand{\Tr}{{\rm Tr}}

\renewcommand{\Im}{{\rm Im}}
\renewcommand{\Re}{{\rm Re}}
\newcommand{\p}{\partial}

\newcommand{\ua}{\uparrow}
\newcommand{\da}{\downarrow}
\newcommand{\ra}{\rightarrow}

\newcommand{\al}{\alpha}
\newcommand{\bt}{\beta}
\newcommand{\del}{\delta}
\newcommand{\Del}{\Delta}

\newcommand{\ga}{\gamma}
\newcommand{\Ga}{\Gamma}

\newcommand{\la}{\lambda}
\newcommand{\La}{\Lambda}

\newcommand{\si}{\sigma}

\newcommand{\cP}{ {\cal P} }

\usepackage{color}
\newcommand {\B}{\textcolor {blue}}

\begin{document}

\title{Triplet $f$-wave pairing in SrPtAs}

\author{Wan-Sheng Wang}
\affiliation{National Laboratory of Solid State Microstructures $\&$ School of Physics, Nanjing
University, Nanjing, 210093, China}

\author{Yang Yang}
\affiliation{National Laboratory of Solid State Microstructures $\&$ School of Physics, Nanjing
University, Nanjing, 210093, China}

\author{Qiang-Hua Wang}
\affiliation{National Laboratory of Solid State Microstructures $\&$ School of Physics, Nanjing
University, Nanjing, 210093, China}

\begin{abstract}

We constructed tight-binding models for the new superconductor SrPtAs according to first principle calculations, and by functional renormalization
group we investigated the effect of electron correlations and spin-orbital coupling (SOC) in Cooper pairing. We found that out of the five $d$-orbitals, the
$(d_{xz},d_{yz})$-orbitals are the active ones responsible for superconductivity, and ferromagnetic spin
fluctuations enhanced by the proximity to the van Hove singularity triggers $f$-wave triplet pairing. The superconducting
transition temperature increases as the Fermi level approaches the van Hove singularity until ferromagnetism sets in. Because of SOC, the spin fluctuations have easy-plane
anisotropy, and the $\v d$-vector of the triplet pairing component is pinned along the out-of-plane direction. Experimental perspectives are discussed.

\end{abstract}



\maketitle

\section{Introduction}

Recently, SrPtAs was found to be a superconductor with a transition temperature $T_c \sim 2.4$ K.~\cite{Nishikubo}
This is a pnictide superdoncutor, but with a hexagonal lattice rather than the square lattice
in iron pnictides. The difference in lattice geometry can lead to completely different
electronic ground states. In square lattices, collinear spin magnetic order is generally
realized (except for systems with ring exchanges), as in cuprates and iron pnictides. \cite{cuprates,iron}
However, a hexagonal lattice would lead to spin frustration,
and even to ferromagnetism. \cite{WWS2,Thomale_k} Since dynamic spin fluctuations can trigger unconventional superconductivity (SC),
the difference in lattice geometry is expected to lead to novel SC. Interesting proposals have
been made, e.g., for Na$_x$CoO$_2$ which also possesses a hexaogonal lattice. \cite{WQH_NCO,Thomale_NCO} An even more
profound aspect of SrPtAs is the conducting element Pt is heavy
hence there is a significant atomic spin-orbital coupling (SOC) among the $5d$-orbitals. Such a coupling can
break spin degeneracy (on general momentum points), modify the Fermi surface topology, and therefore modify low energy particle-hole
excitation spectra. As a result, the effect of SOC is an indispensable factor for unconventional pairing.
Moreover, the unit cell of SrPtAs contains two distinct PtAs layers, each of which has no inversion center, but the system has a global
inversion symmetry with respect to the bisecting plane between the two layers.
The lack of local inversion center opens the possibility of singlet-triplet mixing.~\cite{Sigrist1,Fischer, Maruyama}
This is similar to the mixing in systems where even global inversion symmetry is absent. ~\cite{Gorkov, Bauer, Frigeri}
Combined with strong SOC, such superconductors can exhibit enhanced Pauli limiting fields and a non-vanishing spin
susceptibility down to zero temperature. \cite{Gorkov,Frigeri2,Samokhin,Kimura,Mukuda} Therefore SrPtAs provides a new
playground to explore novel properties of centrosymmetric superconductors with possible singlet-triplet mixing.

Previous local density functional (LDA) calculations for SrPtAs ~\cite{Ivanovskii,Youn1} shows the system is quasi-two dimensional.
There are three pairs of spin-split Fermi surfaces due to SOC. Two of them are centered around the zone center, contributing
about $30\%$ of the total density of states (DOS) at the Fermi level. The remaining $70\%$
comes from the third pair of spin-split Fermi surfaces encircling the $K$ and $K'$ points.
A comprehensive symmetry analysis of the band-resolved pairing symmetry reveals that SrPtAs may possess some unconventional superconducting
states, such as the $A_{2u}$ state with a dominant $f$-wave component and the $E_g$ state with a dominant chiral $d$-wave component.\cite{Sigrist1}
Recently, a muon spin-rotation/relaxation ($\mu$SR) measurement for SrPtAs ~\cite{Biswas} suggests time-reversal symmetry breaking (TRSB)
and a nodeless pairing gap. A nuclear magnetic resonance (NMR) experiment ~\cite{ZhenGq} revealed that
the spin-lattice relaxation rate $1/T_1$ shows a Hebel-Slichter peak below $T_c$, but the peak is strongly
suppressed in another NMR experiment. \cite{Klauss}
Therefore the exact pairing symmetry is still unclear, let alone the pairing mechanism.

The situation motivates us to study the paring mechanism and pairing symmetry of SrPtAs at a microscopic level. For this purpose we construct effective tight-binding models according to LDA band structures. The correlation effects are handled by the singular-mode functional renormalization-group (SM-FRG). \cite{WWS1,XYY1,XYY2,WWS2,XYY3,YY1,YY2}
The advantage of FRG is the capability to survey all electronic instabilities at the same time, \cite{Wetterich}and has been applied with great successes in the contexts of cuprates, \cite{Honerkamp} iron-based superconductors, \cite{WangFa} and more recently for topological SC in correlated systems. \cite{XYY1} As compared to the usual patch-FRG, our SM-FRG has the additional advantages that it respects momentum conservation exactly, and is more straightforward to deal with orbital and spin degrees of freedom.

In this paper, we find that out of the five $d$-orbitals, the $(d_{xz},d_{yz})$-orbitals are the active ones responsible for SC, and ferromagnetic spin fluctuations enhanced by the proximity to the van Hove singularity (VHS) triggers $f$-wave triplet pairing. The superconducting transition temperature increases as the Fermi level approaches the VHS until ferromagnetism sets in. Because of SOC, the spin fluctuations have easy-plane anisotropy, and the $\v d$-vector of the triplet pairing component is pinned along the out-of-plane direction. Experimental consequences of the novel SC are discussed.

\begin{figure}
\includegraphics[width=8.5cm]{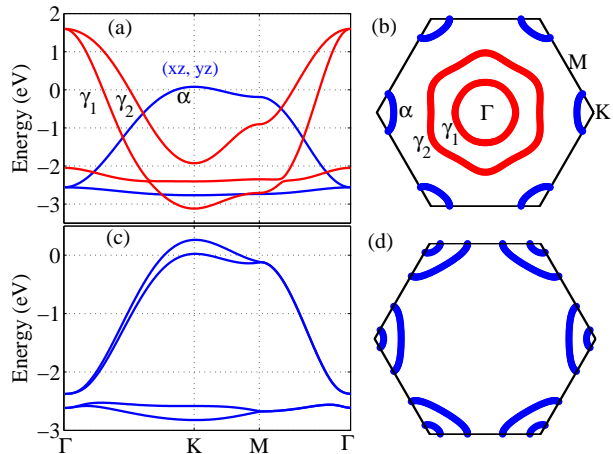}
\caption{(Color online) (a) and (b) are the Spin-degenerate band structure Fermi surfaces for five band
model without SOC. (c) and (d) are spin-split band structure and Fermi surfaces for two-orbital model with SOC.}\label{FS}
\end{figure}

\section{model and method}

Fig.\ref{FS}(a) shows the spin-invariant band structure for SrPtAs obtained by
using the Quantum-ESPRESSO package.~\cite{LDA, LDA2} We then
construct ten maximally localized Wannier functions ~\cite{Wannier} centered at the
two Pt sites in the unit cell, each with five $d$-orbitals
$(d_{3z^2-r^2},d_{xz},d_{yz},d_{x^2-y^2},d_{xy})$.
In agreement with the strong two-dimensionality found in previous LDA calculations, we find
the inter-layer coupling is weak. Thus we shall consider the one layer
model (with one atom per unit cell) for brevity, and will come back to the effect of inter-layer coupling before closing.
The in-plane hopping integrals are presented in Table \ref{TBA}. The orbital-dependent on-site energies are
$(7.577, 8.178, 8.178, 8.787, 8.787)$eV, and finally the Fermi energy is $\mu=9.915$eV.
We notice that although the lattice is hexagonal, the effect of As-atoms lowers the point
group symmetry to $D_{3d}$, and actually to $C_{3v}$ in the effective one-layer model.
The two pockets (labeled as $\ga_1$ and $\ga_2$) around the $\Ga$ point are derived
from $(d_{3z^2-r^2}, d_{x^2-y^2}, d_{xy})$ orbitals, and the pockets around $K$ and $K'$ (labeled as $\al$) are derived from
$(d_{xz}, d_{yz})$ orbitals. The $\al$-pocket is close to the VHS at $M$, and
is expected to be more susceptible to correlation effects than the other pockets.

\begin{table}
\caption{ Hoping integrals $t_{\mu\nu}(\v\Del)$ (in units of eV) where ${\bf \Del}=(\Del_x, \Del_y)$ denotes an in-plane hopping vector, and $(\mu, \nu)$ the orbitals.
Combination of the $C_{3v}$ symmetry and $t_{\mu\nu}({\bf \Del})= t_{\nu\mu}(-{\bf \Del})$ produces  all the in-plain hoppings up to the third neighbors.
Here the five $d$-orbitals $(d_{3z^2-r^2},d_{xz},d_{yz},d_{x^2-y^2},d_{xy})$ are labeled as $(1,2,3,4,5)$ for brevity. }\label{TBA}
\begin{tabular}{|*{5}{c|}}
  \hline
  $(\mu, \nu)\backslash (\Del_x, \Del_y)$ & (1,0) & ( 0 , $\sqrt{3}$) & (0, $-\sqrt{3}$) & (2,0) \\ \hline
  $(1,1)$ &  0.029 &  0.009 &  0.009 &  0.010 \\
  $(1,4)$ &  0.005 &  0     & -0.018 &  0     \\
  $(1,5)$ &  0.022 &  0     &  0     &  0     \\
  $(4,4)$ &  0.158 &  0.043 &  0.043 & -0.022 \\
  $(4,5)$ &  0.135 &  0     &  0     &  0.020 \\
  $(5,5)$ &  0.701 & -0.050 & -0.050 &  0.080 \\ \hline
  $(2,2)$ & -0.456 &  0     &  0     & -0.003 \\
  $(2,3)$ & -0.277 &  0     &  0     & -0.003 \\
  $(3,3)$ &  0.185 & -0.005 & -0.005 &  0.003 \\
  \hline
\end{tabular}
\end{table}

In order to judge the relative importance of the various orbitals, we first perform SM-FRG study of the above
five-orbital model in the absence of SOC. We assume standard local interactions with intra-orbital repulsion
$U$, inter-orbital repulsion $U'$, Hund's rule spin exchange $J$ and pair hopping $J'$, with the details given
in Appendix \ref{RPA_SOC}, and apply the Kanamori relations $U = U'+2J$ and $J=J'$ to reduce the number of
independent parameters. These bare interactions provide the initial values of the running interaction vertices
(versus a decreasing energy scale) in SM-FRG. A general interaction vertex function can be decomposed as
\eq V^{\al, \bt; \ga , \del}_{\v k, \v k', \v q} \ra \sum_m S_m(\v q)
\phi^{\al, \bt}_m(\v k, \v q) [\phi^{\ga, \del}_m(\v k', \v q)]^*, \ee
either in the particle-particle (p-p) or particle-hole (p-h) channel.
Here, $(\al, \bt, \ga, \del)$ are dummy labels for orbital and
spin indices, $\v q$ is the collective momentum, and $\v k$ (or $\v k'$)
is an internal momentum of the Fermion bilinears $\psi^\dag_{\v k + \v q, \al}
\psi^\dag_{-\v k, \bt}$ and $\psi^\dag_{\v k + \v q, \al} \psi_{\v k ,\bt}$
in the p-p and p-h channels, respectively. The fastest growing eigenvalue
$S(\v Q)$ implies an emerging order associated with a collective wave
vector $\v Q$ and eigenfunction (or form factor) $\phi(\v k, \v Q)$. The divergence scale provides an upper limit
of the ordering temperature. In the spin-invariant case one can further resolve spin-density-wave (SDW) and
charge-density-wave (CDW) in the p-h channel. In the p-p channel $\v Q=0$ is always realized
at low energy scale due to the Cooper mechanism. More technical details can be found in
the Appendix \ref{SMFRG} and elsewhere. \cite{WWS1, XYY1}

\section{Results and discussion}

\subsection{Five-orbital model without SOC}

\begin{figure}
\includegraphics[width=8.5cm]{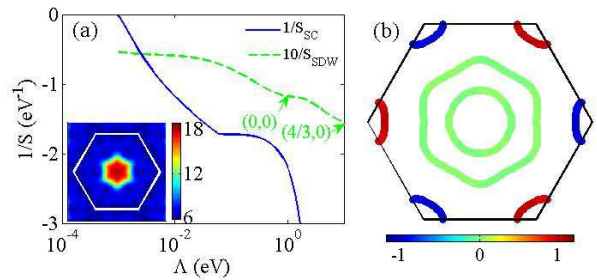}
\caption{(Color online) Results for the five-orbital model without SOC.
(a) FRG flow of $1/S_{SC,SDW}$ versus $\La$. (All interactions are in units of eV.)
The arrows indicate snapshots of the leading $\v q / \pi$ for $S_{SDW}$ during the
flow. The inset shows $|S_{SDW}(\v q)|$ in the momentum space at the final energy scale.
The white hexagonal is the Brillouin zone. (b) The gap function $\Del(\v k)$ on
the Fermi surfaces.}\label{5d}
\end{figure}

We first consider the case in the absence of SOC, where the system is SU(2) invariant.
The FRG flow versus the running energy scale $\La$ (the infrared cutoff of the Matsubara frequency)
for $U=3$eV and $J = U/4$ is shown in Fig.\ref{5d}(a). Since the CDW channel remains weak
at low energy scales, we shall not address it henceforth.
The interaction in the SDW channel, $S_{SDW}$, is enhanced in the intermediate stage and levels
off at low-energy scales. The associated collective momentum $\v Q$ evolves
from $\v Q = (4/3, 0) \pi$ (and its symmetry images) due to high-energy particle-hole excitations between states around $\Ga$ and $K$.
It however settles down at $\v Q = 0$.  The inset of Fig.\ref{5d}(a)
shows $S_{SDW}(\v q)$ versus $\v q$ at the final stage of the flow. A broad peak around $\v q=0$ is apparent.
The form factor $\phi_{SDW}$ turns out to be dominated by site-wise spins from the $(d_{xz},d_{yz})$ orbitals,
in accordance to the VHS near the Fermi level in the $\al$ band.
The strong ferromagnetic fluctuations here is also consistent with the magnetic solutions by LDA. \cite{Youn1}
Attractive pairing interactions, $S_{SC}$ (for $\v Q=0$), is enhanced significantly as $S_{SDW}$ grows (in magnitude).
The cusp in the evolution of $S_{SC}$ is a level crossing of (or change of pairing symmetry in) the leading pairing function $\phi_{SC}(\v k)$.
Eventually $S_{SC}$ diverges so the system will develop SC below the divergence energy scale.
To describe the momentum dependence in the (matrix) function $\phi_{SC}(\v k)$, we introduce the following lattice harmonics
\eqa c_n=\cos (\v k\cdot\v b_n),\ \ s_n=\sin (\v k\cdot \v b_n),\eea where $\v b_{n=1,2,3}$ are the principle translation vectors
$(1, 0),\ ( - 1/2, \sqrt{3}/2)$ and $(-1/2, -\sqrt{3}/2)$, respectively. Up to a global scale, we find
\eqa\phi_{SC}(\v k)\sim (0.68-0.04\sum_n c_n) i\tau_2-0.14i\sum_n s_n\tau_0,\eea where the Pauli $\tau$-matrices operate on $(d_{xz},d_{yz})$ orbitals.
The other elements, including those from the other orbitals, are about two orders
of magnitude smaller than the leading one. The gap function is clearly odd in orbital-momentum space, with $f$-wave symmetry,\cite{note}
thus the spin part must be a triplet by fermion antisymmetry (with three-fold degeneracy because of spin-invariance).
We can project the gap function in the (spin-degenerate) band basis as $ \Del_{\v k} = \< \v k |\phi_{SC}|\v k\>$ where $| \v k \>$ is a Bloch state.
As shown in Fig.\ref{5d}(b), $\Del(\v k)$ is mainly on the $\al$ pocket, and has an $f$-wave symmetry in agreement with the above analysis.
The maximum amplitude of $|\Del(\v k)|$ on the $\ga_1$ and $\ga_2$ pockets are about $400$ times smaller than on the $\al$ pockets.
We conclude that the $\al$-band is active, while the $\ga_1$ and $\ga_2$ bands are passive for SC.
This is an interesting analogue to the situation in Sr$_2$RuO$_4$, \cite{YY1} except that TRS is respected here.\\

\subsection{Effective two-orbital model with SOC}

It is possible to switch on SOC at this stage. However, if all orbitals and form factors are to be kept the numerical demand is beyond our
limit. (The computational complexity is discussed in Appendix A). Instead, we shall consider an effective
two-orbital model with the $(d_{xz},d_{yz})$-orbitals and SOC, guided by the above observation that the
$\al$-band is predominantly active. The validity of such a two-orbital model is justified in Appendix \ref{RPA_SOC}.

The atomic SOC can be written as $H_{SOC} = - \frac{\la}{2}
\sum_i\psi^\dag_i \tau_2 \si_3\psi_i$. Here, the Pauli matrix $\si$ acts on spins. A fit to a relativistic
band-structure calculation \cite{Youn1} yields $\la \sim 0.24$ eV. The spin-split band structure is shown
in Fig. \ref{FS}(c), and the corresponding FS is shown in Fig.\ref{FS}(d) for $\mu = 9.85$ eV.

\begin{figure}
\includegraphics[width=8.5cm]{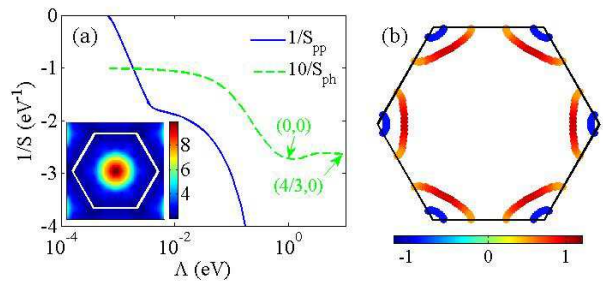}
\caption{(Color online) The results for two-orbital model with SOC.
(a) FRG flow of $1/S_{pp,ph}$ versus $\La$. The arrows indicate snapshots of the leading $\v q / \pi$ for $S_{ph}$ during the
flow. The inset shows $|S_{ph}(\v q)|$ in the momentum space at the final energy scale. The white hexagonal is the Brillouin zone.
(b) The gap function $\Del(\v k)$ on the Fermi surfaces.}\label{2dsoc}
\end{figure}

In the presence of SOC, we apply the SM-FRG extended for spin-resolved fully anti-symmetrized interactions. \cite{XYY1}
The FRG flow for $U = 3$eV and $J = U/4$ is shown in Fig.\ref{2dsoc}(a).
The interaction in the p-h channel, $S_{ph}$, behaves qualitatively similar to $S_{SDW}$ in
Fig.\ref{5d}(a). The inset shows $S_{ph}(\v q)$ versus $\v q$ at the final stage of the flow.
There is a rounded hump at the zone center, but is otherwise similar to the inset of
Fig.\ref{5d}(a). The form factor with the dominant momentum $\v Q=0$ is given by, with twofold degeneracy,
$\phi_{ph}(\v k,\v Q)\sim 0.35\tau_0(\si_x\pm i\si_y).$ (Without SOC the form factor would be three-fold degenerate.)
These form factors are clearly spin-like and are aligned in the plane.
The $\v k$-independence in the leading term means the spin is predominantly site-local.
Thus ferromagnetic spin fluctuations with {\em easy-plane anisotropy}
survive against SOC, although the global magnitude is weakened by roughly a factor of two as compared to the case in the
absence of SOC. Because of the surviving ferromagnetic spin fluctuations, attractive pairing interaction
$S_{pp}$ is also induced and eventually diverges. To reveal the spin and orbital contents explicitly,
we now write the matrix pairing form factor as, $\phi_{pp}(\v k) = (g_{\v k} +
\ga_{\v k})i\si_2$ with singlet and triplet parts $g_{\v k}$ and $\ga_{\v k}$,
respectively. For the case in Fig.\ref{2dsoc}, we find
\eqa g_{\v k} \sim &&  -(0.06 + 0.04\sum_n c_n)\tau_0- 0.1\sum_n s_n \tau_2 \nn
                   &&  + 0.09 [(c_3-c_2)\tau_1+ (2 c_1-c_2-c_3)\tau_3/\sqrt{3}],\\
\ga_{\v k} \sim  && -0.34\sum_n s_n \tau_0\si_3 + (0.03 - 0.17\sum_n c_n)\tau_2 \si_3.\eea
A few remarks are in order. First, the singlet part $g_{\v k}$
transforms as $s$-wave,\cite{note} but the amplitude is relatively small. The dominant
part is the $\tau_{0,2} \si_3$ terms in $\ga_k$ describing $f$-wave triplet
pairing. The fact that these terms are triplets is because the $\tau_0$-term ($\tau_2$-term) is odd in $\v k$ (in the orbital space).
The triplet pairing is clearly triggered by ferromagnetic fluctuations.
The singlet and triplet components mix due to the lack of local
inversion symmetry as well as SOC, and we observe that they transform identically upon
\emph{joint spin-lattice rotations}. One can dub such a pairing as $s^*$-wave according to Ref.\cite{YY2}.
Second, TRS is respected in the above pairing function.
Thus we can project it in the band basis as follows, \eqa  \Del_{\v k} = \< \v k |
\phi_{pp} (\v k) (| - \v k \>)^* = \< \v k | g_{\v k} + \ga_{\v k} | \v k \>,\eea
where $| \v k \>$ is a Bloch state and $| -\v k \> = i\si_2 K | \v k \>$ is the time reversal of $| \v k \>$.
Since $\ga_\v k$ transforms similarly to SOC that splits the bands, it causes a sign change of $\Del(\v k)$ across spin-split bands,
as shown in Fig.\ref{2dsoc}(b) (color scale). However, the pairing gap is nodeless on each spin-split Fermi pockets.
The $s$-wave like sign structure versus rotations is consistent with the previous analysis in the spin-orbital basis.

\begin{figure}
\includegraphics[width=8.5cm]{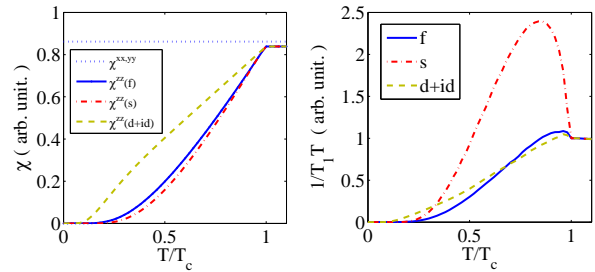}
\caption{(Color online) (a) Spin susceptibility $\chi^{xx,yy}$(blue dotted line) and $\chi^{zz}$(blue solid line) versus temperature for the
 $f$-wave derived from SMFRG, and $\chi^{zz}$ for $s$-wave(red dash-dotted line) and chiral
$d_{x^2-y^2} + i d_{xy}$-wave (green dashed line). The change in $\chi^{xx,yy}$ versus temperature and gap function is almost invisible for the SOC scale
 $\la = 0.24 eV$. (b) The spin lattice relaxation rate $1/T_1T$ versus temperature for the $f$-wave derived from SMFRG
 (blue solid line), and for $s$-wave (red dash-dotted line) and chiral $d_{x^2-y^2} + i d_{xy}$-wave (green dashed line),
 with Dynes factor $\eta = 0.01 T_c$. }\label{chi}
\end{figure}

We emphasize that the above $\ga_\v k$ corresponds to a triplet $\v d$-vector along the $z$-axis, pinned by SOC. This means the total spin of the
Cooper pair along $z$-axis is zero, consistent with the easy-plane spin fluctuations mentioned above,
and implies that in the SC state the out-of-plane spin susceptibility is suppressed, while the in-plane one can survive. This is exactly the case
shown in Fig.\ref{chi}(a), following from a mean field theory calculation using the pairing interaction derived from SM-FRG (see
Appendix \ref{MF_SC}). We also show in Fig.\ref{chi}(a) the spin susceptibilities if the gap function is given by
a fully gapped singlet $s$-wave (red dash-dotted line) and the chiral $d_{x^2-y^2} + i d_{xy}$-wave (green dashed line)
suggested elsewhere. ~\cite{Sigrist1, Fischer_d} We find that the change in $\chi^{xx,yy}$ versus temperature
is almost invisible because of the large SOC scale $\la = 0.24eV$.\cite{notice} On the other hand, $\chi^{zz}$ drops
exponentially right below half of $T_c$ for our $f$-wave and the $s$-wave gaps.  In contrast, it is quasi-linear down to $T_c/4$ for the chiral $d_{x^2-y^2} + i d_{xy}$-wave gap (which is small on the inner $\al$-pocket, and in fact vanishes at the $K$ and $K'$ points).
In Ref.\cite{ZhenGq}, the Knight shift result is well fitted by an $s$-wave gap, but it may also be well fitted by the $f$-wave gap according to the above results.
Fig.\ref{chi}(b) shows the nuclear spin-lattice relaxation rate
$1/T_1T$ versus temperature using the pairing function derived from SMFRG (blue solid line), or given by the $s$-wave (red
dash-dotted line) and chiral $d_{x^2-y^2} + i d_{xy}$-wave (green dashed line) gap. (See Appendix \ref{MF_SC} for technical details.) In the calculation we used a Dynes factor $\eta = 0.01 T_c$ to account for quasiparticle relaxation. We find that $1/T_1T$ has a small peak below $T_c$ for our $f$-wave pairing, while there is a strong peak for the $s$-wave pairing. The peak is minute for chiral $d_{x^2-y^2} + i d_{xy}$-wave pairing. Experimentally, a Hebel-Slichter peak is found in Ref.~\cite{ZhenGq}, but is barely visible in Ref.~\cite{Klauss}.

\begin{figure}
\includegraphics[width=6cm]{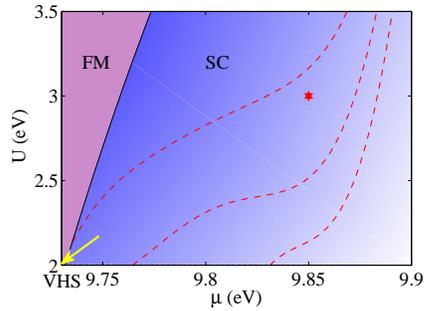}
\caption{(Color online) A schematic phase diagram in $(U,\mu)$ space with
$J=U/4$. The hexagram indicates the case discussed in the text, and the dashed lines indicate equal-value contours for the SC critical scale $\La_c$,
which changes from $0.01$meV, $0.1$meV to $1$meV from right to left. The ferromagnetic order sets in near the VHS (highlighted by
the arrow).}\label{pd}
\end{figure}

We have performed systematic calculations by varying the  bare interaction
parameters and the doping level (or the Fermi level). The results are summarized as a schematic
phase diagram in Fig.\ref{pd}. The pairing scale $\La_c$ increases with hole doping (or decreasing Fermi energy), until the ferromagnetic phase is
approached in the immediate vicinity of the VHS, and in a large regime $\La_c\sim 0.1-1$meV,
of the same order of the experimental $T_c$. We have chosen a fixed ratio $J/U=1/4$ here,
but the results are qualitatively robust down to $J/U=1/12$.
However, if we set $U=U'$ and $J=0$ (up to a small $J'$), we find $d$-wave pairing as in Ref.\cite{Fischer_d}. (A direct comparison to our results is not applicable since the `local' interaction in Ref.\cite{Fischer_d}, defined in the band basis, is not necessarily local in the orbital basis due to orbital hybridizations. \B{Moreover, a pseudo-spin SU(2) symmetry is assumed there but is apparently not respected by the more natural interactions defined in the orbital basis.})\\

Finally we discuss the influence of inter-layer coupling. The inter-plane pairing turns out to be negligible,
and the only effect in the double-layer model is that the singlet component $g_{\v k}$ changes sign
form one layer to the other. Therefore the global symmetry of the gap
function becomes the $A_{2u}$ representation of the $D_{3d}$ group, consistent with the symmetry analysis in Ref.\cite{Sigrist1}. In addition, we find
the spin correlation is ferromagnetic within the plane, but is antiferromagnetic across the plane, in agreement to the LDA calculation.\cite{Youn1}\\

\section{Summary}

We investigated the pairing mechanism in the newly discovered SrPtAs superconductor. We find that out of the five $d$-orbitals, the $d_{xz,yz}$ orbitals are active for superconductivity, and the triplet pairing is driven by FM-like spin fluctuations in SrPtAs. We remark that the pairing function given here respects TRS in the bulk. However, TRSB at grain boundaries is possible and was actually argued as one of the possibilities in the $\mu$SR experiment.\cite{Biswas}
Another possibility is the strong ferromagnetic fluctuations at low energies behave as instantaneous magnetic moments to fast probes
and can cause muon spin relaxations either. Finally, we did not consider the electron-phonon coupling, given the unconventional pairing
revealed in the $\mu$SR experiment. While we can not rule out the role of electron-phonon coupling in driving the superconductivity, our results
do provide a clear picture as to what would correlation leads to if it were the main driving force.\\

\acknowledgments{We thank M. Sigrist for communications, and  C. Platt, W. Hanke and R. Thomale for communications and previous collaborations.
The project was supported by NSFC (under grant No.10974086 and No.11023002) and the Ministry of Science and
Technology of China (under grant No.2011CBA00108 and 2011CB922101). The numerical calculations were performed in the High Performance Computing Center of Nanjing University.}

\appendix
\section{The SM-FRG Method}\label{SMFRG}

The technical details of SM-FRG have appeared in parts (for cases with or without SOC) in Ref.\cite{WWS1,XYY1,WWS2}, and will be rewritten here for self-completeness.
The idea of FRG \cite{Wetterich} is to perform continuous perturbation theory in terms of the change of the phase space. Starting from a high energy window, one obtains the one-particle-irreducible (1PI) vertex functions, and asks how
they change if the infrared limit of the energy window is lowered infinitesimally. This process is repeated versus a running
energy scale, {\em i.e.}, the infrared cutoff $\La$, resulting in a flow of the 1PI vertex functions. The $\La$-dependent vertex
functions provide an effective description of the system at the energy scale $\La$. A diverging four-point vertex function
implies an instability of the normal state toward an emerging order, and the critical scale $\La_c$ is an estimate of
the upper limit of the ordering temperature.

\begin{figure}
\includegraphics[width=8.5cm]{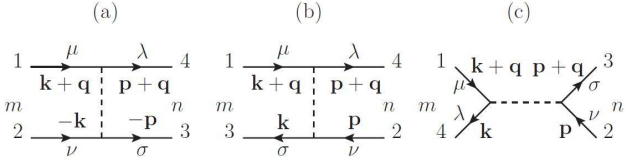}
\caption{A generic 4-point vertex  $\Ga^{1234}$ is rearranged into
$P$-, $C$-, and $D$-channels in (a)-(c), respectively. Here $\v
k,\v q,\v p$ are momenta, $\mu,\nu,\si,\la$ are spin indices, and
$m,n$ denote the basis functions.}\label{pcd}
\end{figure}

Consider a generic four-point vertex function $\Ga$
in the interaction $\psi^\dagger_{\v k_1} \psi^\dagger_{\v k_2} (-\Ga^{1234}_{\v k_1,\v k_2,\v k_3,\v k_4}) \psi_{\v k_3} \psi_{\v k_4}$. The minus sign before $\Ga$ is a convention for later convenience. The labels $1$, $2$, $3$, and $4$ represent orbital-spin indices, and the momentum conservation requires $\v k_1+\v k_2=\v k_3+\v k_4$.
Figs.\ref{pcd} (a)-(c) are rearrangements of $\Ga$ in the pairing (P), crossing (C), and direct (D) channels,
each with a collective momentum $\v q$. The dependence on the other two momenta can be decomposed as
\eqa && \Ga_{\v k+\v q,-\v k,-\v p,\v p+\v q}^{1234}\ra
\sum_{mn}f_m(\v k,1,2)P_{mn}(\v q)f_n^*(\v p,4,3),\nn && \Ga_{\v k+\v q,\v
p,\v k,\v p+\v q}^{1234}\ra \sum_{mn}f_m(\v k,1,3)C_{mn}(\v q)f_n^*(\v
p,4,2),\nn&& \Ga_{\v k+\v q,\v p,\v p+\v q,\v k}^{1234}\ra \sum_{mn}f_m(\v
k,1,4)D_{mn}(\v q)f_n^*(\v p,3,2).\label{projection} \eea
Here $\{f_m\}$ is a set of orthonormal basis functions of the internal momentum $\v k$ (or $\v p$) and a pair of orbital-spin labels. For brevity we shall suppress the
orbital-spin labels in $f_m$ unless indicated otherwise. The momentum dependence in $f_m$ is given by \eq f_m(\v k) = \sum_{\v r} f_m(\v r) \exp( - i \v k \cdot \v r) , \ee
where $f_m(\v r)$ may be chosen to transform according to an irreducible representation of the underlying point group $G$ (which is $C_{3v}$ in the main text),
and $\v r$ is a bond vector connecting the fermion bilinear, e.g., the two $\psi$'s ( or two $\psi^\dag$'s) in Fig.\ref{pcd}(a), or one
$\psi$ and one $\psi^\dag$ in (b) and (c). We notice the decoupling in each channel respects momentum conservation exactly,
since three and only three independent momenta are accessed. On the other hand, if the basis functions form a complete set in momentum, spin and orbital spaces, the above decomposition is exact in each channel, and $P$, $C$ and $D$ are simply different aliases of $\Ga$.

\begin{figure}
\includegraphics[width=5cm]{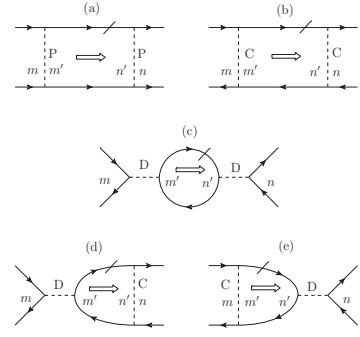}
\caption{One loop diagrams contributing to the flow of the the
4-point vertex function in the pairing channel (a), crossing
channel (b), and direct channel (c)-(e). Here $m,m'n,n'$ denote
basis functions, while the momentum and orbital indices are left
implicit. The open arrows indicate the flow of the collective
momentum. The slashed lines are single-scale fermion propagators.
The slash can be placed on either internal lines associated with
the loop.}\label{oneloop}
\end{figure}

In the absence of SOC, spin conservation enables us to set the spin labels $\mu = \la$ and
$\nu = \si$ in Fig.\ref{pcd}, and in fact they can be suppressed completely.
Under this convention, the one-loop contributions to the flow of the 1PI vertex functions are
shown in Fig.\ref{oneloop}, where (a) and (b) are flows in the pairing and crossing channel,
and (c)-(e) in the direct channel. We denote such contributions as, in matrix form,
\eqa && \p P/\p \La = P \chi'_{pp} P,\nn && \p C/\p\La = C \chi'_{ph} C, \nn && \p D/\p\La = (C-D)
\chi_{ph}' D + D \chi_{ph}'(C-D), \label{f1} \eea
where the collective momentum $\v q$ is left implicit for brevity, and
\eqa
(\chi'_{pp})_{mn} = && \frac{\p}{\p\La}\int\frac{d\w_n}{2\pi}\int\frac{d^2\v
p}{S_{BZ}}f_m^*(\v p)G(\v p+\v q,i\w_n) \nn && \times G(-\v p,-i\w_n)f_n(\v
p)\theta(|\w_n|-\La) \nn = && -\frac{1}{2\pi}\int\frac{d^2\v
p}{S_{BZ}}f_m^*(\v p)G(\v p+\v q,i\La) \nn && \times G(-\v p,-i\La)f_n^*(\v p)\ \
+(\La\ra -\La),\nn
(\chi'_{ph})_{mn} = &&\frac{\p}{\p\La}\int\frac{d\w_n}{2\pi}\int\frac{d^2\v
p}{S_{BZ}}f_m^*(\v p)G(\v p+\v q,i\w_n) \nn && \times G(\v p,i\w_n)f_n(\v
p)\theta(|\w_n|-\La) \nn = && -\frac{1}{2\pi}\int\frac{d^2\v
p}{S_{BZ}}f_m^*(\v p)G(\v p+\v q,i\La) \nn && \times G(\v p,i\La)f_n(\v p)\ \
+(\La\ra -\La),\label{loop} \eea 
where $G$ is the free fermion Green's function, and $S_{BZ}$ is the total
area of the Brillouin zone. Here $\La > 0$ is the infrared cutoff of the Matsubara frequency
$|\w_n|$. As in usual FRG implementation, the self-energy correction and frequency dependence
of the vertex function are ignored.

We observe that $\p P$, $\p C$, and $\p D$ contribute independently
to the full change $d\Ga$, which should be re-interpreted as $dP$, $dC$, and $dD$ in the respective channels. This can
be formally written as \eq d K / d \La = \p K/\p\La + \sum_{K'\neq K} \h \cP_{KK'} [\p K' / \p \La], \label{f2} \ee
for $ K = P, C$ and $D$. Here $\h \cP_{KK'}$ is a projection operator via Eqs.\ref{projection}: it brings the vertex in the $K'$-channel into the form of
the generic $\Ga$, which is subsequently decomposed into the $K$-channel.
In Eq.\ref{f2} the projected terms are overlaps among the three different channels.
It is those terms that allow pairing to be induced by virtual particle-hole scattering processes.
We remark that without the mutual overlap, the flow would correspond to ladder approximations in separate channels.
By taking care of the channel overlap, the full flow in each channel is a faithful representation
of the flow of $\Ga$ if the decomposition in each line of Eq.\ref{projection} is exact.
Clearly FRG treats all channels on equal footing, and the reliability goes
far beyond the scope of simple ladder approximation that overestimates a particular channel while ignoring the others.

At each energy scale $\La$, the effective interaction in the superconducting (SC), spin-density wave (SDW), and charge-density wave (CDW)
channels are given by $V_{SC} = - P$, $V_{SDW} = C$, and $V_{CDW} = C - 2 D$, respectively.
To see the leading instability, we perform singular-value decomposition at each collective momentum $\v q$, for $V=V_{SC/SDW/CDW}$,
\eq V_{mn}(\v q) \ra \sum_i S_i(\v q)\xi_i(m) \eta_i(n), \label{svd}\ee where $S_i(\v q)$ is the
singular value of the $i$-th singular mode, $\xi_i$ and $\eta_i$ are the right and left eigenvectors of $V(\v q)$,
respectively. We fix the phase of the eigenvectors by requiring
$\Re[\sum_m \xi_i(m) \eta_i(m)]>0$ so that $S_i < 0$ corresponds to an attractive mode. In the pairing channel $\v q=\v Q_{SC} = 0 $ addresses
the Cooper instability. In the SDW/CDW channel, the potential ordering wavevector $\v q=\v Q$
is chosen where $S(\v q)$ is maximally attractive. An eigen mode is associated with a
form factor via Eq.\ref{projection}, \eqa \phi_i^{\al,\bt}(\v k)=\sum_m \xi_i(m) f_m(\v k,\al,\bt).\label{phi}\eea
Here we make it explicit that $(\al,\bt)$ is a pair of orbital labels associated with the fermion bilinear in the respective channel.
In terms of such form factors, we can rewrite the interaction vertex as, for a given $\v q$,
\eq V^{\al, \bt; \ga , \del}_{\v k, \v k', \v q} = \sum_i S_i(\v q)
\phi^{\al, \bt}_i(\v k) [\phi^{\ga, \del}_i(\v k')]^*.\ee  The real-space counter part of the form factor,
\eqa \phi_i^{\al,\bt}(\v r)=\sum_m \xi_i(m) f^{\al,\bt}_m(\v r),\eea determines the real-space structure of a candidate order parameter.
For example, in the SC channel the most attractive mode $\phi_{SC}(\v r)$ describes pairing on bond $\v r$, and in the SDW/CDW channel $\phi_{SDW/CDW}(\v r)$
describes spin/charge order on bond $\v r$. Thus both site-local ($\v r=0$) and bond-centered ($\v r\neq 0$) order parameters (and their combinations) in
any channel can be captured. Notice that since $\xi$ evolves during the FRG flow, so does $\phi$. The FRG automatically determines
the most attractive mode with the best form factor. {\em We call such an FRG scheme as the singular-mode FRG (SM-FRG).}\\

\begin{figure}
\includegraphics[width=8cm]{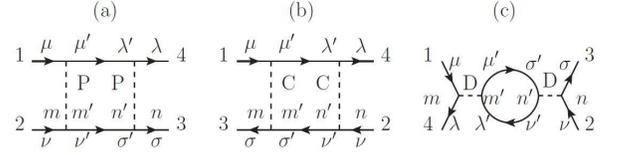}
\caption{One loop diagrams contributing to the flow of the the
fully antisymmetrized 4-point vertex function in the pairing channel (a), crossing
channel (b), and direct channel (c), respectively. Here $m,m'n,n'$ denote
form factors, and $\mu,\nu,\si,\la$ denote spin and orbital indices. }\label{oneloop_soc}
\end{figure}

In the presence of SOC, the spins are not conserved during fermion propagation, and we need to associate a pair of spin indices in $f_m$.
We also need to antisymmetrize the vertex functions explicitly so that the running vertices satisfy fermion antisymmetry. In this case, the matrices $C$
and $D$ are not independent, since $D = -C$.
In the following, $D$ is used for bookkeeping purposes. Figs.\ref{oneloop_soc}(a)-(c) show the one-loop contributions to the flow of
the 1PI vertex functions, with
\eqa \p P / \p \La &=& P \chi'_{pp} P / 2 ,  \nn
     \p C / \p \La &=& C \chi'_{ph} C , \nn
     \p D / \p \La &=& - D  \chi'_{ph} D , \label{flow_soc}\eea
where $\chi'_{pp}$ and $\chi'_{ph}$ are formally identical to that in Eq.\ref{loop}. As in the spin-conserved case,
the full flow equations are given by Eq.\ref{f2}. We notice that the theory reduces to the previous case if spin-invariance is assumed
in the starting hamiltonian.

The effective interaction in the particle-particle (pp) and particle-hole (ph) channels are given
by $V_{pp} = - P/2$ and $V_{ph} = C$, respectively. By singular-value decomposition as
in Eq.\ref{svd} we determine the leading instability in the pp and ph channels.
The (matrix) form factor $\phi^{\al,\bt}(\v k)$ can be constructed for leading eigen modes
as in Eq.\ref{phi}, except that here $\al$ and $\bt$ include spin labels also.\\

A few remarks are in order. First, an emerging collective mode is always associated with a short-range order parameter. For example, the conventional $s$-wave
pairing in the BCS model is local in real space (since the pairing function is independent of momentum), the conventional spin ordered phase is associated with
site-local spins. In cuprates, the $d$-wave pairing occurs primarily on nearest-neighbor bonds, and the $s_\pm$-wave pairing in iron pnictides occurs on
bonds up to the second neighbors. These examples show that in practice it is sufficient to limit $\v r$ in the basis function $f_m(\v r)$ within a given range
in order to capture the leading ordering tendencies. We emphasize that the truncation for $\v r$ does not limit
the collective momentum $\v q$ (or equivalently the setback distance between two fermion bilinears). This is important for us to address the thermodynamic limit.
In our calculations we choose $\v r$ up to the second neighbors, and we checked that longer bonds
do not change the results qualitatively (and even quantitatively). Second, the number of basis functions $f_m$ is $N=N_\v rN_{o}^2N_{s}^2$ where $N_\v r$ is the number of $\v r$'s used for $f_m(\v r)$, $N_{o/s}$ the number of orbitals/spins. In the spin-invariant system the spin label does not enter the flow equations explicitly, so effectively $N_s=1$, while in the presence of SOC we take full account of spins, so that $N_s=2$. In any case the computational complexity in the loop integrations scales as $N^4$, and hence scales as $N_o^8N_s^8$. The quick increase of the complexity versus the number of orbitals is the main computational difficulty in the SM-FRG.
Finally, we remark that our SM-FRG works in the orbital-spin basis. So all relevant bands are taken into account. But the result can be easily transformed into the band basis by simple unitary transformations.\\

\section{Random phase approximation for the five-orbital model with SOC}\label{RPA_SOC}

\begin{figure}
\includegraphics[width=8cm]{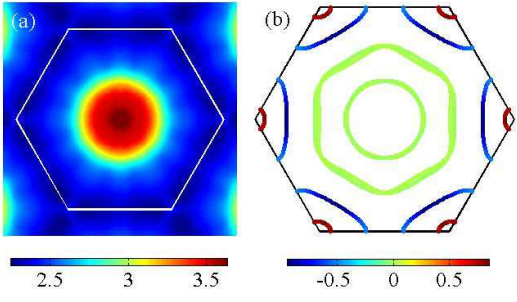}
\caption{The PRA results for $U=1 eV$, $J=U/4$. (a) The leading eigenvalue of
the matrix susceptibility $\chi(\v q)$ as a function of $\v q$ in the Brillouin zone. (b) The leading gap function on the
Fermi surfaces. }\label{rpa}
\end{figure}

As discussed in the above, the computational complexity of the SM-FRG scales as $N_\v r^4 N_o^8 N_s^8$. This forbids us from including all of the five orbitals while retaining all lattice form factors up to the second neighbor bonds (so that $N_\v r=13$). In order to judge the relative importance of the five orbitals in the presence of SOC, we compromise to keep the onsite form factor only (so that $N_\v r=1$), and ignore the overlaps between the pp and ph channels. This enables analytical solution to the flow equations, and the result is identical to that from the standard ladder approximation in the respective channels, except for the infrared energy cutoff. In the ph channel, it is usually referred to as the random phase approximation (RPA). We should remark that in the present setting, there is no divergence in $P$ for repulsive bare interactions, but divergence in $C$ (or in the ph channel) is likely for the RPA scheme overestimates favorable instabilities. We avoid such a divergence by choosing a suitably low energy scale $\La$, at which we transform $-P_\La/2$ and $C_\La-C_\infty$ into the band basis. (The subtraction of the initial value $C_\infty$ is necessary to remove double counting. The subscript $\La$ will be henceforth suppressed for brevity.) We rewrite all of them in the form of pair scattering. They contribute independently to $V_{SC}(\v k,n;\v k',n')$, the pairing interaction for a $(\v k',-\v k')$ electron pair on the $n'$-band to be scattered into a $(\v k,-\v k)$ pair on the $n$-band. The collection of contributions here can be understood as accounting for the channel overlaps once and forever. The pairing interaction is then substituted into the linearized Eliashberg equation,\begin{widetext} \eqa -\sum_{n'}\int\frac{d^2\v k'}{S_{BZ}}V_{SC}(\v k,n;\v k',n')T\sum_{\w_n} G_{n'}(\v k',i\w_n)G_{n'}(-\v k',-i\w_n)\phi_{n'}(\v k')=\la\phi_n(\v k) \eea \end{widetext} to get the leading eigenmode of Cooper pairing with the largest eigenvalue $\la$. Here $T\sim \La$ is the temperature, $G_n(\v k,i\w_n)$ is the Green's function in the $n$-band, and $\w_n$ is the Matsubara frequency. The procedure described here is equivalent to that in the conventional RPA scheme.\cite{LJX_RPA, Scalapino_RPA}

To proceed, we need to specify the atomic SOC involving all of the five $d$-orbitals, $H_{SOC}=-\la / 2 \sum_i \psi_i^\dag \v L \cdot \v \si \psi_i$.
Here, $\v \si$ are pauli matrices. $\v L=(L_x, L_y, L_z)$ is the angular momentum, with the following {\em nonzero} matrix elements in the basis $\psi^t=(d_{3z^2-r^2},d_{xz},d_{yz},d_{x^2-y^2},d_{xy})$,
\eqa L_x^{25} &=& - L_x^{52}= - L_x^{34}= L_x^{43} = i ; \nn
     L_x^{13} &=& - L_x^{31} = \sqrt{3} i ; \nn
     L_y^{35} &=& - L_y^{53}=  L_y^{24}= - L_y^{42} = - i ; \nn
     L_y^{12} &=& - L_y^{21} = - \sqrt{3} i ; \nn
     L_z^{23} &=& - L_z^{32}= -i ; \nn
     L_z^{45} &=& - L_z^{54} = - 2 i. \eea
A fit to a relativistic band-structure calculation yields $\la \sim 0.24 eV$. On the other hand, the local interaction Hamiltonian is given by
\eqa
 H_{I} && = U \sum_{i\alpha} n_{i\alpha\ua}n_{i\alpha\da}
     +U' \sum_{i,\alpha>\beta} n_{i,\alpha}n_{i,\beta} \nn   && +J\sum_{i,\al>\bt,\si\si'}c^{\dag}_{i\alpha\sigma}c_{i\beta\sigma}
    c^{\dag}_{i\beta\sigma'}c_{i\alpha\sigma'}\nn
    && +J'\sum_{i,\alpha\neq \beta}c^{\dag}_{i\alpha\ua}c^{\dag}_{i\alpha\da}
    c_{i\beta\da}c_{i\beta\ua} \eea
with intra-orbital repulsion $U$, inter-orbital repulsion $U'$, Hund's rule spin exchange $J$ and
pair hopping term $J'$. We apply the Kanamori relations $U = U' + 2J$ and $J = J'$ to reduce the
number of independent parameters. We set $U=1 eV$ and $J=U/4$ here for illustration.

Fig.\ref{rpa}(a) shows the leading eigenvalue of the RPA-enhanced ph-channel susceptibility $\chi(\v q)$ (a matrix in the spin-orbital basis). It is peaked around the zone center, consistent with the strong ferromagnetic spin fluctuations found in the effective two-orbital model in Sec.III(B). Fig. \ref{rpa}(b) shows the gap function on the Fermi surfaces for the leading attractive eigenmode of $V_{SC}(\v k,n;\v k',n')$, obtained along the line described above.
The gap function on the $\al$-pockets around $K$ and $K'$  is closely similar to that in the effective two-orbital model in Sec.III(B). Moreover,
the pairing amplitude on the $\ga_{1,2}$ pockets is about $1/40$ of that on the $\al$-pockets. This ratio is larger than that for the five-orbital model without SOC in Sec.III(A), apparently arising from the inter-band proximity effect caused by the SOC. Yet the ratio is still small enough for us to identify the $\al$ pockets as the active bands. Since the latter are dominated by the $d_{xz/yz}$-orbitals, the effective two-orbital model in Sec.III(B) serves as a useful minimal model.

\section{Mean field calculations in the superconducting phase}\label{MF_SC}

If the pp channel is leading, the effective low-energy Hamiltonian is given by
\eq H = H_0 + \frac{V_{eff}}{N} \sum_{\v k, \v k'}B^\dag_{\v k} B_{\v k'} \ee
where $V_{eff}<0$ is the pairing interaction, $N$ is the number of lattice sites, and $B^\dag_\v k$ is the pairing operator
\eqa B^\dag_\v k=\Psi^\dag_{\v k}\phi_{SC}(\v k)(\Psi^\dag_{-\v k})^T,\eea
with the form factor $\phi_{SC}(\v k)$ determined by SM-FRG. Here $\Psi^\dag$ is a spinor creation field
for all orbital/spin degrees of freedom. The mean field hamiltonian can be written as
\eq H_{MF} = H_0 + \sum_{\v k} (\Del B^\dag_{\v k} + {\rm h.c.}), \label{MFH} \ee
subject to the self-consistent condition \eqa  \Del = \frac{V_{eff}}{N} \sum_{\v k} \< B_{\v k} \>.\eea
In the calculation, we choose $V_{eff}$ so that the mean field $T_c$ is close to the FRG divergence scale.

The uniform spin susceptibility $\chi^{\al\al} = \chi^{\al\al} ( \v q \ra 0) $ is given by:
\eq \chi^{\al\al} (\v q) = -\frac{T}{N} \sum_{\v k,\w_n} \Tr \left[G(\v k,i\w_n) \ga^\al G(\v k+\v q,i\w_n) \ga^\al\right], \ee
where $G$ and $\ga^\al$ are the Green's function and spin vertex (of polarity $\al$) in the Nambu space.
The trace is taken in the spin-orbital-Nambu space.

The spin-lattice relaxation rate $1/T_1T$ is associated with the low frequency dynamics of local spins, and is given by
\eqa \frac{1}{T_1T} = &-& \lim_{\nu \to 0} \sum_{\al \bt} \frac{g_{\al \bt}}{\nu} \Im \Tr[\ga_\al G_{loc}(i\w_n+i\nu_n) \nn
  &\times& \ga_\bt G_{loc}(i\w_n)]| _{i\nu_n \ra \nu+i0^+} \nn
= &-& \la^2 \sum_\al \int d \w \frac{\p f}{\p \w} \Tr [\ga_\al A(\w) \ga_\al A(\w)] .  \eea
Here $g_{\al \bt} = \la^2 \delta_{\al \bt}$ is taken as the hyperfine coupling matrix element, $f$ is the
Fermi function, and $G_{loc}$ is the local Green's function, which can be expanded as
\eq G_{loc}(i\w_n) =\frac{1}{N} \sum_{\v k,m} \frac{| \v k ,m \> \< \v k, m |}{i\w_n - E_{\v k,m}} = \int d\w \frac{A(\w)}{i\w_n-\w}, \ee
where $|\v k,m\>$ is the eigenstate, and we defined a local spectral matrix
\eqa A(\w) &=& \frac{1}{N}\sum_{\v k m} | \v k,m\>\<\v k,m| \delta(\w - E_{\v k,m}) \nn
&\ra & \frac{1}{N}\sum_{\v k m}\frac{\eta}{\pi}\frac{ | \v k,m\>\<\v k,m|} {(\w- E_{\v k,m})^2 + \eta^2 }. \eea
In the last line we approximate the delta-function by a Lorentzian with a Dynes factor $\eta$.


\begin{references}
\bibitem{notice} We remark that the in-plane $\chi^{xx,yy}$ hardly change across $T_c$ here, since the SOC is present and is much larger than
the mean field pairing gap. For consistency we checked that in the limit of $\la\ra 0$,  $\chi^{xx,yy}$ for singlet pairing become identical to $\chi^{zz}$.
\end{references}
\end{document}